\documentclass[preprint,prd,noshowpacs,nofootinbib]{revtex4}
\usepackage{amssymb}
\usepackage{amsmath}

\begin{document}

\title{Global  versus Local -- Mach's Principle versus the Equivalence Principle
\footnote{``Honorable Mention" award in the Gravity Research Foundation 2016 Awards for Essays on Gravitation. \\
Corresponding Author: D. Singleton}}

\author{Douglas Singleton}
\email{dougs@csufresno.edu}
\affiliation{Physics Department, CSU Fresno, Fresno, CA 93740 USA
and \\
ICTP South American Institute for Fundamental Research,
UNESP - Univ. Estadual Paulista
Rua Dr. Bento T. Ferraz 271, 01140-070, S{\~a}o Paulo, SP, Brasi}

\author{Steve Wilburn}
\email{stevejwilburn@gmail.com}
\affiliation{Twin Prime Inc., Redwood City, CA 94063 USA}

\date{\today}

\begin{abstract}
The equivalence principle is the conceptual basis for general relativity.
In contrast Mach's principle, although said to have been influential on Einstein in his formulation of 
general relativity, has not been shown to be central to the structure of general relativity. In this 
essay we suggest that the quantum effects of Hawking and Unruh radiation are a manifestation of a {\it thermal} 
Mach's principle, where the local thermodynamic properties of the system are determined by the non-local structure of the
quantum fields which determine the vacuum of a given spacetime. By comparing Hawking and Unruh temperatures for the 
same local acceleration we find a violation of the Einstein elevator version of the equivalence principle, which vanishes 
in the limit that the horizon is approached.       
\end{abstract}

\maketitle

{\noindent {\large {\bf The Equivalence Principle, Mach's Principle and Gravity}}}

One of the ``coincidences" in physics is the equivalence of gravitational and inertial mass. 
In constructing a theory of gravity Einstein elevated this equivalence to the central principle 
which forms the conceptual foundation of general relativity \cite{einstein}. Being based 
on a physical principle distinguishes general relativity from quantum mechanics. One could 
argue that the recent and loosely formulated ``holographic principle" \cite{hologram} provides 
such a principle for quantum mechanics which might allow one to construct a 
quantum theory of gravity. 

There are different ways to formulate the equivalence principle. First on comparing the inertial 
mass from Newton's 2$^{nd}$ Law ({\it i.e.} ${\bf F} = m_i {\bf a}$) with the mass that appears in 
Newton's Law of Gravitation ({\it i.e. ${\bf F} = \frac{G m_g M_g}{r^2} {\bf {\hat r}}$}) one finds, that 
to the limits experimentally tested, the two masses are equivalent $m_i = m_g$. 

A second formulation of the equivalence principle -- and the one used chiefly in this essay -- 
is in terms of Einstein's elevator \cite{einstein}. If an observer placed inside a small 
enough enclosure, such as an elevator, feels their feet pressed to the bottom
of the elevator, this observer can not tell if the elevator is at rest on the surface of some gravitating planet,
with a local gravitation field ${\bf g}$, or if the elevator is accelerating through empty spacetime with acceleration
${\bf a}={\bf g}$. The condition ``small enough" means that tidal forces, which {\it do} differentiate 
between a gravitational field and acceleration, can be ignored. This locality condition gives general relativity a local character. 

In counterpoint Mach's principle, which is said to have inspired Einstein, does not share the same central 
role as the equivalence principle in the foundations of general relativity. This is partly due to the fact that 
Mach's principle is vaguely defined. In this essay we will take Mach's principle to have the following meaning:
``The inertial properties of an object are determined by the energy-momentum throughout all space" \cite{wheeler}.
This statement makes it clear that Mach's principle is, at least partly, a global principle -- the local 
inertial properties of a particle are determined by the global distribution of energy-momentum throughout the entire 
space. Given this tension between the global nature of Mach's principle versus the local character of the  
equivalence principle, it might not be surprising that only one of these principles has found a firm footing
in the foundations of general relativity. However, by comparing the temperatures of Hawking and Unruh radiation 
we will find support for a {\it thermal} Mach's principle and a violation of the Einstein elevator
version of the equivalence principle.         
\\

{\noindent {\large {\bf Hawking and Unruh radiation}}}

Hawking radiation \cite{hawking} is the radiation emitted by a black hole of mass, $M$. It occurs 
as a consequence of placing quantum fields in the gravitational background of a black hole. An observer who stays at a 
{\it fixed} distance, $r$, from a black hole of mass, $M$, will measure radiation with thermal spectrum and
a temperature given by (here $c=1$ and $k_B =1$ but Newton's constant $G$ and Planck's constant $\hbar$ are kept) \cite{wilburn}
$$
T_{Hawking} = \frac{\hbar }{8 \pi G M \sqrt{1- \frac{2GM}{r}}} ~.
$$
The static observer we have in mind is exactly the one described in section 4.2 of \cite{barbado}. 
Also as described in \cite{barbado} the above temperature is obtained by the observer only after some
time. This time is short -- for a 1.5 solar mass black hole and with the observer fixed at a distance of 12 km
the time is $\approx 10^{-4}$ s.  Normally, the Hawking temperature $T_{Hawking}$ is quoted for an observer a 
large distance from the black hole ({\it i.e.} $r \rightarrow \infty$) so the factor $\sqrt{1- \frac{2GM}{r}}$ goes to 1. 
This observer, who measures the above Hawking temperature and is at a fixed distance, $r > 2GM$, outside a 
black hole of mass, $M$, will experience a local acceleration \cite{wilburn}
$$
a_{BH} =  \frac{1}{\sqrt{1- \frac{2 G M}{r}}} \left(\frac{G M}{r^2} \right) ~.
$$
Now by the Einstein elevator version of the equivalence principle an observer, accelerating through Minkowski spacetime, should also 
measure thermal radiation \footnote{In order for the Einstein elevator version of the equivalence principle to apply one needs the size of the 
elevator and all measuring devices to be small compared to variation of the gravitational field over the size to the elevator. Here
we have in mind using an Unruh-DeWitt (UD) detector \cite{bd} to measure the local temperature of the given spacetime. 
An UD detector is a point-like, 2-state quantum system whose excitation of the upper energy state can be used to measure the 
temperature of a given spacetime. Since the UD detector is point-like one is always in the regime where the equivalence principle applies.} otherwise the accelerating observer could immediately tell the difference between being in the gravitational field of 
a black hole versus being in an accelerating frame. 
Soon after Hawking's paper on black hole radiation it was shown that an accelerating observer
(with an acceleration of $a=|{\bf a}|$) does detect thermal radiation ({\it i.e.} Unruh radiation) with a 
temperature ({\it i.e.} the Unruh temperature) given by \cite{unruh}
$$
T_{Unruh} = \frac{\hbar a}{2 \pi } ~.       
$$
The above Unruh temperature is that measured by the unrealistic case of an observer who has been 
accelerating forever, from $t=-\infty$ to $t=+\infty$. However,
in \cite{merkli} it was shown that an accelerating UD detector will asymptotically approach a Gibbs thermal state 
with the above temperature $T_{Unruh}$ regardless of its initial state. 
Since the accelerated observer measures a temperature, at least qualitatively, there is 
no violation of the equivalence principle. The first observer in the Einstein
elevator {\it fixed} at a distance, $r$, from a black hole will measure both a downward acceleration (given above) 
toward the floor of the elevator and thermal radiation at a temperature $T_{Hawking}$; the second observer in 
the Einstein elevator which is accelerating through Minkowski spacetime will measure the same downward acceleration 
toward the floor of the elevator and also thermal radiation. However, looking at this situation {\it quantitatively} 
uncovers a violation of the equivalence principle {\it except} in the limit as the observer approaches the event horizon \cite{wilburn}.

In order to compare an Einstein elevator at rest in the gravitational field of a black hole
with an Einstein elevator accelerating through empty space, one sets the acceleration of the 
elevator to the same acceleration experienced by the observer in the gravitational field of a 
black hole $a = a_{BH}$. Using this acceleration, $a_{BH}$, in $T_{Unruh}$ gives 
$$
T_{Unruh} = \frac{\hbar}{2 \pi \sqrt{1- \frac{2 G M}{r}}}\left( \frac{G M}{r^2}\right) ~.
$$
For locations outside the event horizon of the black hole ({\it i.e.} $r> 2GM$) one can easily check that 
$T_{Hawking} > T_{Unruh}$. For example, suppose that the observer is fixed over a black hole of mass $M$ at a distance 
$r= 4GM$ ({\it i.e.} at twice the Schwarzschild radius). The observer would measure a local acceleration of
$a_{4GM}=\frac{\sqrt{2}}{16 G M}$. At this point the observer would not know if they are in a gravitational field 
produced by a mass $M$ or accelerating through empty space. They would only measure $a_{4GM}$.
Now inserting $a_{4GM}$ into the Unruh temperature gives 
$T_{Unruh}= \frac{\hbar a_{4GM}}{2 \pi}=\frac{\sqrt{2} \hbar}{32 \pi G M}$. On the other hand the Hawking temperature for this set up 
is $T_{Hawking}=\frac{\sqrt{2} \hbar}{8 \pi G M} = \frac{2 \hbar a_{4GM}}{\pi} = 4 T_{Unruh}$ {\it i.e.} four times
the Unruh temperature. Finally by going to the UD detector and 
reading off the temperature the observer could immediately tell if they are accelerating through empty space 
(in which case the UD detector would read the first temperature, $T_{Unruh}$) or if they are fixed in a 
gravitational field (in which case the UD detector would read the second, higher temperature,
$T_{Hawking}$). This works for any $r>2GM$ ({\it i.e.} outside the event horizon). 

In summary the equivalence principle violating thought experiment for an observer equipped with an accelerometer
and a thermometer \footnote{As in reference \cite{wilburn} the ``thermometer" we have in mind is a point-like
UD detector so that one is always in the regime where the equivalence principle applies.} is as follows: (i) Measure the local acceleration, 
$a=|{\bf a}|$. (ii) Insert this local acceleration into the expression for $T_{Unruh} = \hbar a / 2 \pi $. (iii) Measure the temperature. If this temperature is higher than that calculated in step (ii) then one is in the gravitational field of a black hole
and not in an accelerating frame. 
 
In limit $r \rightarrow 2GM$ 
$$
T_{Hawking} \rightarrow \frac{\hbar }{8 \pi G M \sqrt{1- \frac{2GM}{r}}} = T_{Unruh} ~.
$$
Thus, in the near horizon region the equivalence principle is restored -- at the horizon both 
temperatures diverge to the same infinite value due to the factor $\frac{1}{\sqrt{1- \frac{2GM}{r}}}$. This divergence 
is as expected since for an observer {\it fixed} just above the horizon the local acceleration and Hawking temperature 
both diverge. This violation of the equivalence principle is reminiscent of the firewall puzzle 
given in \cite{amps} where one encounters a ``firewall" of high energy particles at the black hole horizon. However,
in \cite{amps} this divergence at the horizon signals a potential breakdown of the equivalence principle  
whereas here the divergence at the horizon leads to a restoration of the equivalence principle in the sense that
$T_{Hawking} = T_{Unruh}$ in the limit $r \rightarrow 2 G M$. Another example where the equivalence principle may be
violated when quantum mechanics and general relativity are combined, are neutrino oscillations in a gravitational background
\cite{gasperini} \cite{ahluwalia} \cite{mureika}. 

In the above discussion we have focused on the temperature measured by the observer.
For observers away from the horizon the actual spectrum is modified from pure Planckian, 
$\propto \frac{1}{exp[\hbar \omega/ T] - 1}$, to the form
$\propto \frac{\gamma (\omega)}{exp[\hbar \omega/ T] - 1}$ where $\gamma (\omega)$ is the greybody factor
which is a function of the frequency of the emitted
quanta. Although the spectrum is distorted by $\gamma (\omega)$ the temperature which can be extracted from
this spectrum is not altered by the greybody factor. A similar point was made in \cite{matas} -- that if one
looks at the full response rate of the UD detector one also gets a violation of the equivalence principle
when one is away from the horizon. It is for this reason that we have focused here on the temperature --
it is unchanged by the greybody factor. \\

{\noindent {\large {\bf Thermal Mach's Principle}}}

One way to understand the reason for the violation of the equivalence principle given above is that the 
equivalence principle is fundamentally a local principle -- a gravitational field and acceleration are only 
equivalent in a small enough spacetime region. In contrast Hawking and Unruh radiation depend on how
one defines the quantum vacuum \cite{bd} which is a non-local construction. Briefly, to describe this construction,
suppose one has a scalar field, $\phi$, which obeys the wave equation $\Box _g \phi =0$ where the subscript $g$ indicates the type of spacetime
one has ({\it e.g.} Minkowski, Rindler, Schwarzschild, {\it etc.}). The solution to $\Box _g \phi =0$ has a set
of mode solutions $u_k (x)$ where the subscripts $k$ are the momenta of the particular field mode. 
One can then expand a general field $\phi$ in terms of these modes as
$\phi (x) = \sum _k [a_k u_k(x) + a_k ^\dagger u_k ^* (x) ]$ where $a_k , a_k ^\dagger$ are annihilation/creation
operators of quanta of momenta $k$. The operators $a_k , a_k ^\dagger$ define the vacuum state.
Note also that the modes, $u_k (x)$, are non-local in extent. In any case whether one has Minkowski vacuum, Rindler vacuum,
Boulware vacuum, Hartle-Hawking vacuum, the vacuum state is a non-local construction. Since Hawking and Unruh radiation depend
on non-local vacuum states it is not surprising that one finds these effects violate the equivalence
principle which is local.

The preceding discussion suggests a {\it thermal} Mach's principle: the system's local thermal properties such as the 
temperature, radiation spectrum, and entropy depend on the non-local nature of the quantum vacuum of the spacetime.
This is seen to be similar to the statement quoted earlier from \cite{wheeler} where the local inertial properties of the
system are determined by the global distribution of the energy-momentum in the spacetime.

One could then argue that the above proposed thermal Mach's principle sheds light on the original Mach's principle where 
the local inertial properties are determined by the global energy-mass distribution. The agreement of the temperatures, 
$T_{Hawking} = T_{Unruh}$, 
in the horizon limit, $r \rightarrow 2 G M$, fits nicely with the proposed holographic models of \cite{verlinde} and \cite{gogber} where 
inertia arises from the entropy of the system. Combining the thermal Mach's principle with the idea from these holographic models,
that inertia comes from thermodynamic quantities such as entropy, may suggest the original, inertial version of Mach's principle
as a conceptual basis for a quantum theory of gravity. \\

{\noindent {\bf Acknowledgment:}} DS is supported by a 2015-2016 Fulbright Grant and by grant $\Phi.0755$ by the Ministry of Education and Science of Kazakhstan.

{\noindent {\bf Note added:}} After this essay was completed a paper appeared \cite{ng} which showed a violation of the EP
via the use of an UD detector in the more amazing/provocative case of a detector in Minkowski spacetime versus a detector
inside a hollow shell.

\end{document}